\documentstyle[preprint,aps]{revtex}
\tightenlines
\begin{document}
\draft
\flushbottom
\preprint{\vbox{
\hbox{IFT-P.035/95}
\hbox{hep-ph/9507448}
}}
\title{ The heavy top quark and right-handed currents}
\author{Vicente Pleitez }
\address{
Instituto de F\'\i sica Te\'orica\\
Universidade Estadual Paulista\\
Rua Pamplona, 145\\ 
01405-900-- S\~ao Paulo, SP\\
Brazil}
\maketitle
\begin{abstract}
We consider a modification of the standard electroweak model with
the third quark generation and the $\tau$-lepton in vector
representations of $SU(2)\otimes U(1)_Y$ electroweak symmetry.  
This is a new way to implement right-handed 
currents which are controlled by the usual
Fermi constant, $G_F$, the weak mixing angle, $\sin\theta_W$, 
and also by the right-handed mixing matrices which survive when the Lagrangian 
density is written in terms of the mass eigenstates. In this case there are 
also new $CP$ violation phases.
\end{abstract}

\section{Introduction}
\label{sec:intro}
The standard model of the elementary particles based on the 
gauge symmetry $SU(3)_C~\otimes SU(2)_L~\otimes U(1)_Y$~\cite{sm} is in 
agreement with almost all experimental data~\cite{pdg}. Data which are 
not easily accommodated in this model are not definitive yet~\cite{new2}.
The electroweak part of this model is built in such a way that all 
charged currents are purely left-handed and parity is maximally violated 
in weak processes. Thus, the electroweak standard model (ESM) is a complete 
chiral model {\it i.e.}, the left-handed fermions transform in a different way 
than the right-handed ones do. Moreover, all generations transform in the same 
way and this is the reason why parity is maximally violated in all of them.

Since the left-handedness of the bottom quark has 
not been experimentally tested yet with the same precision than that obtained
in the measurements for the other fermions, it has been pointed out, recently, 
that in the context of $SU(2)_L\otimes SU(2)_R\otimes U(1)$ electroweak
models, this quark may decay through, in the extreme case, 
purely right-handed~\cite{lr}. A more realistic situation is to assume that the 
third generation has right-handed currents which are not dominant in the weak
decay of the $b$-quark. This issue will be 
set by studying the decay of polarized $\Lambda_b$, in the $b\to s\gamma$ 
decay~\cite{japa} or, as it will be argued below, in some processes at 
the $Z^0$-peak energy.

In fact, in literature when right-handed currents among the known generations
are considered it is mainly done in the context of models involving right-handed
gauge bosons, like in the left-right symmetric models mentioned 
above~\cite{lrm}. In 
this case the effect of right-handed currents (and any correction to the
results obtained with the ESM) are controlled chiefly by the two parameters:
the mass of the new gauge bosons in the form of the parameter 
$\delta=(M_L/M_R)^2$, with $M_L\approx M_W$ and the mixing angle 
$\zeta$ among left- and right-handed vector bosons~\cite{kuz}.

Here we will show that it is possible to have right-handed currents controlled 
by the usual parameters $G_F$, $\sin\theta_W$, and also by the right-handed 
mixing matrices which survive in the Lagrangian density. These right-handed 
mixing angles and phases may be important only for the third generation, 
keeping consistency with the phenomenology of the first two generations. 
We recall that, although parity violation has been tested mainly with the 
first and second generations~\cite{gb} it may not necessarily happen with the 
third generation. 
Hence, even in the context of a $SU(2)\otimes U(1)$ model we can have
right-handed currents 
because new parameters, that are not constrained in the context of the ESM, 
survive in the Lagrangian density when one of the generations transforms in a 
vector way under the electroweak symmetry. 
We call this type of models {\it quasi}-chiral models.

Before going on this new way to implement right-handed currents, we first
emphasize some features of the mass matrices for quarks in the context of 
the ESM. In this model, in order to
diagonalize the quark mass matrices it is necessary to introduce unitary 
matrices $V^{U}_{L,R}$ and $V^D_{L,R}$ in both chiral (left- and right 
fields) and charge ($u$- and $d$-like) sectors.  
There is also automatic flavor conservation in both, neutral 
Higgs-fermion and neutral vector-fermion interactions. 
This implies that the right-handed mixing matrices $V^U_R, \,V^D_R$ 
disappear when the Lagrangian density is 
written in terms of the quark mass eigenstates $U=(u,c,t,...)^T$ and 
$D=(d,s,b,...)^T$. 
Only the left-handed mixing matrices $V^U_L$ and $V^D_L$ remain
in the charged current coupled to the $W^\pm$: $\bar{U}_LV_{\rm CKM}\gamma^\mu
D_L\,W^-_\mu$ where $V_{\rm CKM}=V^{U\dagger}_LV^D_L$ denotes the Cabibbo-
Kobayashi-Maskawa mixing matrix. 
Thus, no trace of the matrices $V^{U,D}_{L,R}$ separately is left in the 
Lagrangian density of the ESM. 
However, as we will show below, this argument is not valid when at least 
one generation is assigned to a vector representation of $SU(2)\otimes U(1)_Y$.

Summarizing, the main goal of this work is to show that: if we consider the 
electroweak interactions in a quasi-chiral model, with the left- and
right-handed components of the third 
generation transforming as doublets under $SU(2)$, the right-handed 
mixing matrices survive after the diagonalization of the mass matrices.
Another interesting consequence is that the physical phases in the 
Lagrangian density are more than one, that is, new sources of CP violation 
arise too in this context (see Sec.~\ref{sec:con}).

This work is organized as follows. In Sec.~\ref{sec:model} we present the 
modification of the standard electroweak model in which the third 
generation belongs to a vector representation of $SU(2)\otimes U(1)_Y$. The  
Yukawa interactions are given in Sec.~\ref{sec:yuk} while gauge interactions 
are considered in Sec.~\ref{sec:inte}. Our conclusions and some phenomenological 
considerations appear in the last section.

\section{A modified $SU(2)\otimes U(1)_Y$ model}
\label{sec:model}
As we said before, in this work we consider a version of the ESM  in which the
third family is in a vector representation of $SU(2)\otimes U(1)_Y$. 
We define the electric charge operator as usual, $Q/e=I_3+Y/2$, so that
the quark sector in this model is 
\begin{equation}
Q_{iL}=\left( 
\begin{array}{cc}
u_i \\ d_i\end{array}
\right)_L\sim \left({\bf2},\frac{1}{3}\right);\quad
u_{iR}\sim\left({\bf1},\frac{4}{3}\right),
\quad d_{iR}\sim\left({\bf1},-\frac{2}{3}\right), \quad i=1,2;
\label{1}   
\end{equation}
as in the ESM, but with the third quark generation transforming as
\begin{equation}
Q_{3L}=\left(
\begin{array}{cc}
u_3 \\ d_3\end{array}
\right)_L,\quad Q_{3R}=\left(
\begin{array}{cc}
u_3 \\ d_3\end{array}
\right)_R\sim\left({\bf2},\frac{1}{3}\right). 
\label{2}     
\end{equation}

The Higgs boson sector consists of the usual doublet $\varphi=
(\varphi^+, \varphi^0)^T$, a singlet $\phi\sim({\bf1},0)$ and a triplet:
$H\sim({\bf3},0)$
\begin{equation}
\vec{H}\cdot\vec{\tau}=
\left(
\begin{array}{cc}
H^0_T & \sqrt{2}H^- \\
\sqrt{2}H^+ & -H^0_T
\end{array}
\right),
\label{e3}
\end{equation}
with $\vec{\tau}$ being the Pauli matrices. The shifted neutral fields 
are defined as $\varphi^0=(1/\sqrt2)(v_D+H^0_1),\;\phi=(1/\sqrt2)(v_S+H^0_2)$ 
and $H^0_T=(1/\sqrt2)(v_T+H_3^0)$, and we will assume $v_S> v_D\gg v_T$. 
We can always choose $v_D$ real by an appropriate $SU(2)$ transformation.
The triplet does not contribute to the mass of the $Z$ but it does
to the $W$ mass. Hence we have new contributions to the $\rho$ parameter which 
measures the strength of the neutral current interaction relative to the 
charged current interaction and that at tree level in this model is defined as 
$\rho\equiv M_W/M_Z\cos\theta_W=1+4(v_T/v_D)^2$. The 
global fit for this ratio is $1.0012\pm0.0013\pm0.0018$~\cite{pdg}. It means 
that $v_T\approx9.7$ GeV in order to be consistent with data, say at 2 standard
deviations. At higher order there will be other contributions to $\Delta\rho$ 
as in the ESM but these have to be computed in the context of the present 
model. The value of $v_S$, of course, is not constrained by present experimental 
data.

Next, let us consider leptons. To put the tau lepton and its neutrino
in a vector representation of $SU(2)$ might be inconsistent 
with phenomenology. However, we can introduce a fourth lepton family and 
neutral singlets in order to have
more room for experimental data, especially the $Z$-invisible
width and also for maintaining the model anomaly free. 
Then we have the left-handed fields transforming as $({\bf2},-1)$
\begin{mathletters}
\label{leptons}                                 
\begin{equation}
\Psi_{eL}=\left(
\begin{array}{c}
\nu_e \\ e^-
\end{array}\right)_L,\;
\Psi_{\mu L}=\left(
\begin{array}{c}
\nu_\mu \\ \mu^-
\end{array}\right)_L,\;
\Psi_{\tau L}=\left(
\begin{array}{c}
\nu_\tau \\ \tau^-
\end{array}\right)_L,\;
\Psi_{T L}=\left(
\begin{array}{c}
\nu_T \\ T^-
\end{array}\right)_L;
\label{leptons1}                              
\end{equation}
right-handed components transforming also as $({\bf2},-1)$
\begin{equation}
\Psi_{\tau R}=\left(
\begin{array}{c}
N_1 \\ \tau^-
\end{array}\right)_R,
\Psi_{TR}=\left(
\begin{array}{c}
N_2 \\ T^-
\end{array}\right)_R.
\label{leptons2}
\end{equation}
\end{mathletters}
Finally, the singlets $N_{1L},N_{2L}\sim({\bf1},0)$ and $e_R,\mu_R\sim
({\bf1},-2)$. 

It is interesting to notice that it is possible to embed 
this model in a gauge symmetry $SU(3)_C\otimes SU(3)_L\otimes U(1)_N$ as in
the model proposed some years ago by Georgi and Pais~\cite{gp,vp}. In order to
do this we must introduce extra quarks with left- and right- handed components
transforming as singlets under $SU(2)$: two with charge $-1/3$, 
say $s'$ and $b'$, and one with charge $2/3$, say $t'$. 
This isosinglets have effects on $Z\to b\bar b,c\bar c$~\cite{ema} that 
eventually must be considered in the phenomenology of the model.
Here we will not consider this possibility. 

\section{Yukawa interactions}
\label{sec:yuk}
Let us now consider the generation of the fermion masses. In the quark sector,
the Yukawa interactions are
\begin{eqnarray}
-{\cal L}_Y&=&\sum_{i,j}\,\bar Q_{iL}\,[A_{ij}\,u_{jR}\tilde\varphi+
B_{ij}\,d_{jR}\varphi]+\sum_i\,\bar Q_{3L}\,[a_i\,u_{iR}\,\tilde\varphi
+b_i\,d_{iR}\,\varphi]\nonumber \\ & &\mbox{}
+\sum_\alpha\,h_\alpha\,\bar Q_{\alpha L}\,Q_{3R}\,\phi
+\sum_{\alpha,i,j} h'_\alpha\bar Q_{\alpha iL}\,Q_{3jR}\,
(\vec{H}\cdot\vec{\tau})_{ij}
+H.c.,
\label{4}           
\end{eqnarray}
where $i,j=1,2$ and $\alpha=1,2,3$. All coupling constants 
$A$'s, $B$'s, $a$'s, $b$'s $h$'s and $h'$'s are, in principle, complex numbers. 

In the quark sector we have from Eq.~(\ref{4}) the mass matrices
\begin{equation}
M^U=\frac{v_S}{\sqrt2}\,\left(
\begin{array}{ccc}
 A_{11}r & A_{12}r & h_1 \\
A_{21} r& A_{22}r & h_2  \\
a_1r & a_2 r & h_3+h'_3r'
\end{array}
\right),\qquad
M^D=\frac{v_S}{\sqrt2}\,\left(
\begin{array}{ccc}
B_{11}r & B_{12}r & h_1 \\
B_{21}r & B_{22}r & h_2 \\
b_1r & b_2 r & h_3-h'_3r'
\end{array}
\right),
\label{6}             
\end{equation}   
with the notation $r=v_D/v_S$, $r'=v_T/v_S$. We have assumed, for the sake of 
simplicity, a discrete symmetry allowing only the third generation to couple
with the triplet.
Notice that, since there are three scalars contributing to the
mass matrices, there are flavor changing neutral currents (FCNC). 

The scalar triplet breaks strongly the isospin symmetry. For simplicity we will 
assume that $h_3,h'_3>0$.  We see from Eq.~(\ref{6}) that
$m_t\approx (v_S/\sqrt2)(h_3+h'_3r')$ and $m_b\approx (v_S/\sqrt2)(h_3-h'_3r')$,
so that $m_t-m_b=\sqrt{2}v_Th'_3$ and $m_t+m_b=\sqrt{2}v_Sh_3$. 
Thus, we have that if $v_T=9.7$ GeV, 
$m_t=179$ GeV and $m_b=4.5$ GeV, $h_3\approx 0.32$ ($h_3^2/4\pi\approx 0.008$) 
and $h_3'\approx13$ ($h'^2_3/4\pi\approx 13$) and it appears that at least the 
third generation interacts strongly with the Higgs triplet. 
This sort of scalar could be implemented by a dynamical symmetry breaking 
where the Higgs bosons appear as condensates of the fermions present in the 
model, but we will not speculate about this issue here.

In general, these mass matrices can be diagonalized using unitary 
matrices $V^U_{L,R}$ and $V^D_{L,R}$ defined as follows:  
\begin{equation}
U'_L=V^U_LU_L,\quad U'_R=V^U_RU_R;\quad D'_L=V^D_LD_L,\quad
D'_R=V^D_RD_R,
\label{7}               
\end{equation}
where $U'=u_1,u_2,u_3$ and $D'=d_1,d_2,d_3$ are symmetry
eigenstates, while $U=u,c,t$ and $D=d,s,b$ are mass eigenstates. 
Mass matrices in Eq.~(\ref{6}) are general enough to generate the 
mass spectra $m_t\gg m_c > m_u$; $m_b>m_s>m_d$. 

From Eq.~(\ref{4}) we have the quark-scalar interactions
\begin{eqnarray}
-{\cal L}_Y&=& \bar U_L\left[\frac{\hat{M}^U}{v_D} -\frac{1}{r}\,
V^{U\dagger}_L\Omega V^U_R\right]U_R\,H^0_1
+\bar D_L\left[\frac{\hat{M}^D}{v_D} -\frac{1}{r}\,
V^{D\dagger}_L\Omega V^D_R\right]
D_R\,H^0_1\nonumber \\ & &\mbox{}
+\left[\bar U_LV^{U\dagger}_L\Omega V^U_RU_R+\bar D_L V^{D\dagger}_L
\Omega V^D_R D_R\right]H_2^0
\nonumber \\ & &\mbox{} 
+[\bar{U}_L V^{\dagger U}_L \Omega^+V^U_RU_R+ 
\bar{D}_LV^{\dagger D}_L\Omega^- V^D_R]H^0_3
\nonumber \\ & &\mbox{}
+\sqrt{2}h'_3\left(\bar{D}_L V^{D\dagger }_L\Delta V^U_RU_RH^-
+\bar{U}_L V^{U\dagger }_L \Delta V^D_R D_RH^+\right) + H.c.
\label{yukawa}                       
\end{eqnarray}
where $\hat{M}^U,\hat{M}^D$ are the diagonal mass matrices for the $u$-like 
and $d$-like quarks, respectively, $\Delta={\rm diag}(0,0,1)$ and
\begin{equation}
\Omega=\frac{1}{\sqrt2}\,\left(
\begin{array}{ccc}
0\; & 0\; & h_1  \\
0\; & 0\;  & h_2  \\
0\; & 0\;  & h_3
\end{array}
\right),\quad
\Omega^{\pm}=\frac{1}{\sqrt2}\,\left(
\begin{array}{ccc}
0\; & 0\; &  0\\
0\; & 0\;  &  0\\
0\; & 0\;  &\pm h'_3r'
\end{array}
\right).  
\label{12}                              
\end{equation}   

We see that in the interactions of quarks with the scalar $H_1^0$ the
factor $1/r>1$ appears. However, since the mixing matrices $V^{U\dagger}\Omega
V^U_R$ and $V^{D\dagger}\Omega V^D_R$ 
also appear, these interactions are nor necessarily strong. The 
interactions with the scalar $H^0_3$ are proportional to $h'_3r'$ so, they also
are not strong. On the other hand, the interactions with the charged
scalars $H^\pm$ are proportional
to $h'_3$, hence, these are in principle, strong interactions
depending on the values of the parameters in $V^{D\dagger }_L\Delta V^U_R$
and $ V^{U\dagger }_L \Delta V^D_R$. The matrices $V^{U,D}_{L,R}$ also control
the flavor changing neutral currents in the scalar sector.
Recall too that the fields $H^0_{1,2,3}$ in Eq.~(\ref{yukawa}) are not the 
mass eigenstates yet. 
 
The Yukawa interactions in the lepton sector are 
\begin{eqnarray}
-{\cal L}_l&=&\sum_{a,b}\bar\Psi_{aL}G_{ab}l_{bR}\varphi+
\sum_{a,\alpha}\bar\Psi_{aL}h_{a\alpha}N^c_{\alpha R}\tilde\varphi+
\sum_{m,\alpha}\bar\Psi_{mR}F_{m\alpha}N_{\alpha L}\tilde\varphi
\nonumber \\ & &\mbox{}
+\sum_{a,m}\bar\Psi_{aL}H_{am}\Psi_{mR}\phi
+\frac{1}{2}\,\sum_{\alpha,\beta}
\bar{N}_{\alpha L}\,h'_{\alpha\beta}N^c_{\beta R}\,\phi
+\sum_{m,n;i,j}\bar{\Psi}_{niL}H'_{mn}\Psi_{njR}\, 
(\vec{H}\cdot\vec{\tau})_{ij}+H.c.
\label{yulep}                                 
\end{eqnarray}
where $l_{bR}=e_R,\mu_R$; $G_{ab},h_{a\alpha},H_{am},F_{m\alpha}$ and 
$h'_{\alpha\beta}$ denote arbitrary complex dimensionless constants, 
$a=e,\mu,\tau,T$;  
$m,n=\mu,T$ and $\alpha,\beta=1,2$. From Eq.~(\ref{yulep}) we obtain an 
arbitrary $4\times 4$ mass matrix for charged leptons. 
On the other hand, assuming $G_{ab}v_D, H'_{mn}v_T<H_{am}v_S\ll 
F_{m\alpha}v_D$, the neutral leptons $N_\alpha$ are heavier than the 
charged leptons. In fact, if we also assume 
$h_{a\alpha}v_D<h'_{\alpha\beta}v_S\ll F_{m\alpha}v_D$, 
then the $N_\alpha$'s have its mass matrix  
dominated at tree level by the term $v_D\bar N_{\alpha R}F_{\alpha\beta}
N_{\beta L}$. We have too the mass term $\frac{1}{2}\bar \psi_{\nu_L} 
M \psi^c_{\nu_R}$, where $\psi_{\nu L}=(\nu_{eL},\nu_{\mu L},\nu_{\tau L},
\nu_{TL},N^c_{1L}, N^c_{2L})^T$ and
\begin{equation}
M\approx v_S\left(
\begin{array}{cccccc}
0 & 0 & 0 & 0 & h_{e 1}r     & h_{e2}r     \\
0 & 0 & 0 & 0 & h_{\mu 1}r  & h_{\mu 2}r  \\
0 & 0 & 0 & 0 & h_{\tau 1}r & h_{\tau 2}r \\
0 & 0 & 0 & 0 & h_{T1}r     & h_{T2}r     \\
h_{e 1}r & h_{\mu 1}r & h_{\tau 1} & h_{T 1}r
  & h'_{11} & h'_{12}                                \\
  h_{e 2}r & h_{\mu 2}r & h_{\tau 2}r & h_{T 2}r
  & h'_{21} & h'_{22}
\end{array}
\right).
\label{M}                                        
\end{equation}
This matrix has two eigenvalues equal to zero and four others different from 
zero. It is possible, as in the case of three lepton
doublets and one neutral singlet~\cite{tau3}, to 
fit the $Z^0$ invisible width without assuming a see-saw hierarchy in 
Eq.~(\ref{M}).

\section{Gauge interactions}
\label{sec:inte}
The neutral current interactions of fermions $\psi_i$ can be written as
usual
\begin{eqnarray}
{\cal L}^{NC}&=&-\frac{g}{2c_W}\,\left[
\sum_iL_i\bar\psi_{iL}\gamma^\mu \psi_{iL} +R_i\bar\psi_{iR}\gamma^\mu
\psi_{iR}\right]\,Z_\mu, \nonumber \\ &= &
-\frac{g}{2c_W}\,\sum_i\bar\psi_i\gamma^\mu(g_{Vi}-\gamma^5g_{Ai}) 
\psi_i\,Z_\mu,
\label{nc}                                          
\end{eqnarray}
where $g_{Vi}\equiv\frac{1}{2}(L_i+R_i)$, $g_{Ai}\equiv \frac{1}{2}
(L_i-R_i)$ and $c_W\equiv \cos\theta_W$. Thus, we have (using also 
$s^2_W\equiv \sin^2\theta_W$)
\begin{mathletters}
\label{lravd}                                          
\begin{equation}
L_{d_1}=L_{d_2}=L_{d_3}=-1+\frac{2}{3}\,s^2_W\,\equiv L_D\,;
\label{d1}
\end{equation}
\begin{equation}
R_{d_1}=R_{d_2}=\frac{2}{3}\,s^2_W\equiv R_{D},
\quad R_{d_3}=-1+\frac{2}{3}\,s^2_W\equiv R'_{D};
\label{d2}
\end{equation}
\end{mathletters}
for the charge $-1/3$ sector, and
\begin{mathletters}                                      
\label{lravu}
\begin{equation}
L_{u_1}=L_{u_2}=L_{u_3}=1-\frac{4}{3}\,s^2_W\equiv L_U\,;
\label{u1}
\end{equation}
\begin{equation}
R_{u_1}=R_{u_2}=-\frac{4}{3}\,s^2_W\equiv R_{U},\quad 
R_{u_3}=1-\frac{4}{3}\,s^2_W\equiv R'_{U};
\label{u2}
\end{equation}
\end{mathletters}
for the charge $2/3$ sector. Here $L_D,R_D$ and $L_U,R_U$ denote the
ESM couplings for the respective charge sector.

For the leptons we have
\begin{mathletters}
\label{leplr}                                                
\begin{equation}
L_{\nu_a}=1,\;  L_a=-1+2s^2_W,\;\; a=e,\mu,\tau,T;\quad L_{N_\alpha}=0,
\;\;\alpha=1,2;
\label{l1}
\end{equation}
and
\begin{equation}
R_{\nu_a}=0,\;R_e=R_\mu=2s^2_W,\;R_\tau=R_T=-1+2s^2_W,\; R_{N_\alpha}=1.
\label{l2}
\end{equation}
\end{mathletters}
Notice that the singlets $N_\alpha$ have only right-handed neutral 
currents.

The left-handed neutral currents in Eq.~(\ref{nc}) can be rewritten, 
in the $d$-like sector, as
\begin{equation}
{\cal L}^{NC}_{dL}=-\frac{g}{2c_W}\,L_D\,(\bar d_{1}\;\bar d_{2}\; 
\bar d_{3})_L\gamma^\mu
\left(
\begin{array}{ccc}
1\; &\; 0 & 0\; \\
0\; & 1\; & 0\; \\
0\; & 0\; & 1 \;
\end{array}
\right) \left(
\begin{array}{c}
d_{1} \\ d_{2} \\ d_{3}
\end{array}\right)_L\,Z_\mu,
\label{ncul}                                                    
\end{equation}
and we see that this interaction is diagonal in terms of the symmetry 
eigenstates. On the other hand, the respective right-handed neutral 
currents are 
\begin{equation}
{\cal L}^{NC}_{dR}=-\frac{g}{2c_W}\,
(\bar d_{1}\;\bar d_{2}\; \bar d_{3})_R\gamma^\mu
\left(
\begin{array}{ccc}
\frac{2}{3}s^2_W\;\; & 0 \;\;& 0 \\
0\;\; & \frac{2}{3}s^2_W\;\; & 0 \\
0\;\; & 0\;\; & -1+\frac{2}{3}s^2_W
\end{array}
\right)  \left(
\begin{array}{c}
d_{1} \\ d_{2} \\ d_{3}
\end{array}\right)_R\,Z_\mu.
\label{ncur}                                                
\end{equation} 

Both Eqs.~(\ref{ncul}) and (\ref{ncur}), are given
in terms of the symmetry eigenstates. When the transformations in 
Eq.~(\ref{7}) are done, the left-handed currents in Eq.~(\ref{ncul}) are 
still diagonal, thus the Glashow-Iliopoulos-Maiani (GIM) mechanism~\cite{gim} 
is implemented at tree level in the left-handed currents for the charge $-1/3$ 
sector. The same occurs in the left-handed charge $2/3$ sector.
Notwithstanding, this does not happen in the right-handed currents. Hence,
there are flavor changing right-handed neutral currents (FCRNC) at tree 
level. When we turn to the mass eigenstates,
Eq.~(\ref{ncur}) becomes, using the transformation given in Eq.~(\ref{7}), 
\begin{equation}
{\cal L}^{NC}_{dR}=-\frac{g}{2c_W}\bar D_R\gamma^\mu \left[ R_D{\bf 1}-
V^{D\dagger}_R \Delta V^D_R\right]D_R\,Z_\mu,
\label{n1}                                                        
\end{equation}
where $D=(d,s,b)$ denotes the mass eigenstates, $\Delta= diag(0,0,1)$ as in 
Sec.~\ref{sec:yuk}, and $R_D=(2/3)s^2_W$. We see that the FCRNC involves 
some elements of the matrix $V^D_R$ which do not survive in the standard model 
after the diagonalization of the mass matrix. 

Similarly for the $u$-type quarks we have
\begin{equation}
{\cal L}^{NC}_{uR}=-\frac{g}{2c_W}\bar U_R\gamma^\mu \left[ R_U{\bf 1}-
V^{U\dagger}_R \Delta V^U_R\right]U_R\,Z_\mu,
\label{n2}                                                          
\end{equation}
where $U=(u,c,t)$ and $R_U=-(4/3)s^2_W$.

Concerning the charged currents of the model, besides the usual left-handed 
ones which coincide with the currents of the standard model, 
we have the contribution $\bar u_{3R}\gamma^\mu d_{3R}$ that, in terms
of the mass eigenstates is written as
\begin{equation}
{\cal L}^{CC}_{R}=-\frac{g}{\sqrt2}\,
(\bar u\;\;\bar c\;\; \bar t)_R\,
\gamma^\mu V^{U\dagger}_R\,\Delta
V^D_R\left(
\begin{array}{c}
d \\ s \\ b\end{array}
\right)_R\,W^+_\mu+H.c.,
\label{ru}                                                            
\end{equation}  
and the only relevant parameters could be $\left( V^{U\dagger}_R\Delta V^D_R
\right)_{ub}$  and $\left( V^{U\dagger}_R\Delta V^D_R\right)_{cb}$ 
(other matrix elements can be $\approx0$). Hence, the weak charged currents can
still be predominantly left-handed as in the standard model~\cite{cleo}. 
Notice that in Eq.~(\ref{ru}) both $V^U_R$ and $V^D_R$ do appear.  

These new currents are controlled by the matrix 
elements of $V^{U\dagger}_R$ and $V^D_R$, besides $G_F$ and 
$\sin^2\theta_W$ which also appear in the other standard currents.
Neither the pa\-ra\-me\-ters in Eqs.~(\ref{n1}) and (\ref{n2}) 
nor those in Eq.~(\ref{ru}) appear explicitly in the standard model. As we said 
before, in the last case,
only the combination $V_{\rm CKM}=V^{U\dagger}_LV^D_L$ is left when the 
Lagrangian density is written in terms of the 
mass eigenstates. Hence, in this model it is possible that both, charged and 
neutral right-handed currents, couple mainly to the third family.
(See also Sec.~\ref{sec:con}).

In the leptonic sector we do not have the GIM mechanism at tree level in 
the right-handed neutral currents coupled to the $Z^0$ because the mass matrix 
mixes the $\tau$ lepton with the other charged leptons. So, there are also 
FCRNC effects. In terms of the mass eigenstates $e,\mu,\tau$ and $T$ 
we have the following interactions (in Eq.~(\ref{leptons}) the same notation 
denotes symmetry eigenstates)
\begin{equation}
{\cal L}^{NC}_{lR}=+\frac{g}{2c_W}\,(\bar{e}\;\;\bar{\mu}\;\;
\bar{\tau} \;\; \bar{T})_R\gamma^\mu\,V^{l\dagger}_RY^l_RV^l_R\,\left(
\begin{array}{c}
e\\\mu\\ \tau\\T
\end{array}
\right)_RZ_\mu,
\label{rleptons}                          
\end{equation} 
where $V^l_R$ is the $4\times 4$ right-handed mixing 
matrix in the charged lepton sector and we have introduced
$Y^l_R=\mbox{diag}(2s^2_W,2s^2_W,-1+2s^2_W,-1+2s^2_W)$.
We can write Eq.~(\ref{rleptons}) as follows
\begin{equation}
{\cal L}^{NC}_{lR}=+\frac{g}{2c_W}\,\bar l_R\gamma^\mu
\left[ R_l-V^{l\dagger}_R \Delta^l V^l_R \right]l_R\,Z_\mu,
\label{n3}                                   
\end{equation}
where $R_l=2s^2_W$ and $\Delta^l=diag(0,0,1,1)$. 
We see that after we have diagonalized the mass matrix 
of the charged leptons, the right-handed mixing matrix survives in 
Eq.~(\ref{rleptons}). 

The values of these new mixing parameters can be chosen in 
such a way that the model can be consistent with the measurements of $\tau$ 
asymmetries at LEP. In fact, it has been measured the ratio of vector to 
axial-vector neutral couplings and it has been found consistency with the 
hypothesis of $e-\tau$ universality~\cite{pdg}. 
There are measurements of the tau-neutrino 
helicity~\cite{aa} and Michel parameters~\cite{aam} which confirm a dominant 
left-handed ($V-A$) structure in the charged current for the $\tau$ and its 
neutrino and by which pure vector $(V)$, axial-vector $(A)$ or right-handed 
$(V+A)$ interactions have been ruled out. However, in the present model we 
have $V+A$ interactions with almost arbitrary strength, so this experimental 
data will imply only constraints on some mixing angles in $V^l_R$. We recall 
that in this model there are also flavor changing neutral currents in the 
neutrino sector like in the model with a single neutral singlet~\cite{tau3}. 
Thus at tree level we can 
impose that in Eq.~(\ref{n3})
\begin{equation}
V^{l\dagger}_{Rea} V^l_{Rae}\approx V^{l\dagger}_{R\mu a} 
V^l_{Ra \mu}
\approx V^{l\dagger}_{R\tau a} V^l_{Ra\tau}\approx 0,\quad a=e,\mu,\tau;
\label{vatau}                                   
\end{equation} 
with the other elements of the matrix with $a=T$ to be 
set up by other experimental data. 

The charged currents of the phenomenological neutrinos 
$\nu_a\;(a=e,\mu,\tau,T)$ with all 
charged leptons are purely left-handed, as can be seen from 
Eq.~(\ref{leptons1}). However, when they are written in terms of the mass 
eigenstates, because $N_1$ and $N_2$ have purely right-handed couplings with 
$\tau^-$ and $T^-$ as can be seen from Eq.~(\ref{leptons2}), 
right-handed couplings among neutrino mass eigenstates and charged leptons
will appear.
So, we can write right-handed currents like in Eq.~(\ref{ru}), in the 
leptonic sector. As there are six neutrinos but only four charged leptons, it
is possible to extend the column of the charged leptons with two zeros in such 
a way that the right-handed current is written in terms of $6\times 6$ matrices involving 
both mixing matrices in charged and neutral sector. We will not write these 
interactions explicitly.

\section{Conclusions and phenomenological consequences}
\label{sec:con}
It is not our intention to make here a detailed study about the phenomenology
of the model. We want only to point out some remarkable features.
Up to present all experimental data 
have not been sufficiently to rule out deviation
from the $V-A$ structure. In the context of the $L-R$ symmetric models 
they only put constrains on the parameters
$\delta=(M_L/M_R)^2$ and the mixing angle $\zeta$. For instance, studying 
direct neutron beta decay as it is done in Ref.~\cite{kuz}. In the present model
we have several parameters to be fitted with experimental data. Hence, we have
more freedom than in $L-R$ models.

As we said before, almost all $Z^0$- pole observables 
are in agreement with the standard model predictions~\cite{pdg}. There 
are, however, some of them which are under special consideration at present 
because it seems that they do not agree completely with the 
model's predictions. For example, the heavy quark production ratios
\begin{equation}
\frac{\Gamma(Z^0\to q\bar{q})}{\Gamma(Z^0\to\hbox{hadrons})}\equiv
\frac{\Gamma_q}{\Gamma_h},\quad
\Gamma_h=\sum_q\Gamma_q,
\label{51}                          
\end{equation}
that have been measured for $c$ and $b$ quarks. Until recently, 
$\Gamma_b/\Gamma_h$ and $\Gamma_c/\Gamma_h$ were in conflict with the ESM.
Nowadays, although $Z\to c\bar c$ agrees with the theoretical predictions, 
$Z\to b\bar b$ lies some $1.8\sigma$ above the respective predictions of 
the ESM~\cite{new2}. 

Another observable which is in disagreement with the standard model is 
the forward-backward asymmetry $A^{(0,b)}_{LR}\approx A_eA_b$, with $A_f$ 
defined as
\begin{equation}
A_f\equiv \frac{2g_V^fg_A^f}{(g^{f}_V)^2+(g^{f}_A)^2}=\frac{(L_f)^2-(R_f)^2} 
{(L_f)^2+(R_f)^2}.
\label{ab}
\end{equation}
The SLD data give
\begin{equation}
A_b=0.882\pm0.068(\mbox{stat})\pm 0.047(\mbox{sys}),
\label{absld}
\end{equation}
leading to a world average 
\begin{equation}
A_b=0.867\pm0.022,
\label{abw}
\end{equation}
about 3$\sigma$ below the standard model prediction $A_b=0.935$~\cite{new2}.

These observables $\Gamma_b/\Gamma_h$ and $A^{(0,b)}_{LR}$ depend on the 
effective vector and axial-vector couplings $\bar{g}_{Vi}$ and 
$\bar{g}_{Ai}$, defined in Eq.~(\ref{nc}), which in the ESM scenario 
include radiative corrections.  
In the context of our model $\bar{g}_{Vi}$ and $\bar{g}_{Ai}$ or 
$\bar L$'s and $\bar R$'s refer to the couplings defined in terms of the 
coefficients which incorporate the right-handed neutral currents in 
Eqs.~(\ref{n1}), (\ref{n2}) and radiative corrections too. 
We will use the notation 
$\delta R_{ff'}\equiv\bar R_{ff'}-R_F=2\delta^{ff'}_R$ (or just $2\delta^f_R$, 
if $f=f'$) and we will write 
them only at tree level, $R_F$ denotes the ESM value without radiative 
corrections for all fermions of the same charge sector i.e., $F=U,D$, 
while $f,f'=d,s,b,u$ and $c$. We write explicitly only the diagonal
case $f=f'$:
\begin{mathletters}
\label{vaeffd}                         
\begin{equation}
\delta R_{dd}\equiv \bar R_d-R_D=\left( 
V^{D\dagger}_R \Delta V^D_R\right)_{dd}=\left(V^{D\dagger}_{R}\right)_{bd}
\left(V^D_{R}\right)_{bd}=2\delta g^d_R;
\label{va1}
\end{equation}
\begin{equation}
\delta R_{ss}\equiv \bar R_s-R_D=\left( 
V^{D\dagger}_R \Delta V^D_R\right)_{ss}=\left(V^{D\dagger}_{R}\right)_{bs}
\left(V^D_{R}\right)_{bs}=2\delta g^s_R;
\label{va2}
\end{equation}
\begin{equation}
\delta R_{bb}\equiv \bar R_b-R_D=\left( 
V^{D\dagger}_R \Delta V^D_R\right)_{bb}=2\delta g^b_R=1-2\delta g^d_R
-2\delta g^s_R;
\label{va3}
\end{equation}
\end{mathletters}
for the $d$-like sector, and
\begin{mathletters}
\label{vaeffu}
\begin{equation}
\delta R_{uu}\equiv \bar R_u-R_U=\left( 
V^{U\dagger}_R\Delta V^U_R\right)_{uu}=\left(V^{U\dagger}_{R}\right)_{tu}
\left(V^U_{R}\right)_{tu}=2\delta g^u_R;
\label{va4}
\end{equation}
\begin{equation}
\delta R_{cc}\equiv \bar R_c-R_U=\left( 
V^{U\dagger}_R\Delta V^U_R\right)_{cc}=\left(V^{U\dagger}_{R}\right)_{tc}
\left(V^U_{R}\right)_{tc}=2\delta g^c_R;
\label{va5}
\end{equation}
\end{mathletters}
for the $u$ and $c$ quarks; the $\delta$'s on the right hand side 
denote the respective coefficients in the notation of Ref.~\cite{cpb}. 
The factor 2 arises because of the different parameterization between 
the neutral currents in our Eq.~(\ref{nc}) and Eq.~(2) in Ref.~\cite{cpb}. 
On the other hand, in the lepton sector we have from Eq.~(\ref{rleptons})
\begin{equation}
\delta R_{ll}\equiv \bar R_{l}-R_L=-\left(V^{l\dagger}_R\Delta^l 
V^l_R\right)_{ll}=2\delta ^l_R.
\label{n4}                                         
\end{equation}
where $R_L$ denotes the ESM value for the charged leptons.
 
Using Eq.~(\ref{va3}), we can write Eq.~(\ref{ab}) for $f=b$ as
\begin{equation}
A_b=\frac{L_D^2-(R_D+\delta R_{bb})^2}{L_D^2+(R_D+\delta R_{bb})^2},
\label{new1}
\end{equation}
and if $\delta R_{bb}\approx0.071$ or $-0.380$ we obtain a value for
$A_b$ which is compatible with that in Eq.~(\ref{abw}). Notice however that, 
from Eqs.~(\ref{vaeffd}), this implies that $\delta R_{dd}=\delta R_{ss}
\approx0.35$ or 0.23. 
Only a careful analysis looking for a correlation among all the parameters in 
the model may show if there is some range for these parameters fitting all 
these experimental data. It is possible to fit the $Z\to b\bar b$ width (even 
if this width will coincide in the future with the numerical expectation of the 
standard model) and the $A^{(0,b)}_{LR}$ asymmetry if at the same 
time the light quark and lepton contributions are appropriately modified.
So, even assuming that all nondiagonal right-handed couplings are negligible,
we have a possible correlation among the six parameters: $\delta^b_R,
\delta^c_R$, and $\delta^{UD}$ defined in our context as
\begin{equation}
\delta g^{UD}_R=4\sum_{q=u,d,s}g^q_R\delta g^q_R, 
\label{n5}                                           
\end{equation}
together with $\delta^e_R, \delta^\mu_R$ and $\delta^\tau_R$.  
We also stress that in the present model there are nonstandard $W$ 
couplings, as in Eq.~(\ref{ru}), and FCRNC effects parameterized by $\delta
R_{sb}$, etc. Independent of the anomaly in the $A_b$ asymmetry to persist 
or not, it is necessary to look for direct evidence of the right-handed $b$ 
couplings.

As we said before, it is well known that
the leptonic charged currents are 
consistently left-handed~\cite{aa,aam} and also that the $c$ quark has a 
dominant left-handed structure~\cite{charm}. However, these results put 
constrains only on the allowed values for some elements of 
the matrices $V^{U,D,l}_{L,R}$ entering in the first two generation processes. 
In fact, for the $c$ quark a small mixture of right-handed couplings can be 
accommodated by data. In $c$ meson decays, using the CDHS detector, 
the upper limit of $0.07$ for the relative strength of the 
square of the $c\to s$ right-handed current has been obtained~\cite{charm}. 
The measurement of the lepton decay forward-backward asymmetry in the reaction 
$\bar B\to D^*l^-\nu_l$ confirms that the chirality of the weak interactions 
in $b\to c$ transitions are predominantly left-handed~\cite{cleo}. However, we 
must recall that experimental measurements in $\bar B\to D^*l^-\nu_l$ decay 
assume that the lepton current is left-handed~\cite{cleo}, but in our model 
leptons have both, left- and right-handed currents. (See the discussion in the 
last paragraph of Sec.~\ref{sec:inte}.) More recently, also in $b\to X l\nu_l$ 
transitions, the dominant $(V-A)\times (V-A)$ structure of the weak charged 
currents has been confirmed. The maximal $(V+A)\times (V-A)$ is 
excluded with a significance of 5.7 and 3.7 standard deviations for the $b\to 
Xe\nu_e$ and $b\to X\mu \nu_\mu$ cases, respectively. For the $b\to Xe\nu_e$ 
the $V\times (V-A)$ structure is excluded by more than 3.5 standard deviations 
while for the $b\to X\mu\nu_\mu$ case it is found to be equally consistent with the
$(V-A)\times (V-A)$ and the $V\times (V-A)$ case~\cite{L3}. 

This does not exclude some right-handed current effects, for instance, 
by considering the four-fermion interaction for a semileptonic decay
of $b$ 
\begin{equation}
2\sqrt2 G_F V_{cb}\left[(\bar c_L\gamma^\mu b_L)+
\xi(\bar c_R\gamma^\mu b_R) \right](\bar e_L\gamma^\mu \nu_L),
\label{p1}
\end{equation}
with $\xi=g_R/g_L$, it has been shown that the difference of the value of 
$\vert V_{cb}\vert$ 
extracted from the total inclusive semileptonic decay rate of the $B$
mesons and the amplitude of the exclusive decay $B\to D^*l\nu$ is 
sensitive to an admixture of a right-handed $b\to c$ current.
This can be interpreted in terms of the $\xi$ parameter in Eq.~(\ref{p1}) and 
it has been found that $\xi\approx0.14\pm0.18$~\cite{voloshin}.
In the present model we have interactions like that in Eq.~(\ref{p1})
if we consider Eq.~(\ref{ru}) and the usual left-handed charged currents
in the quark sector. However we also have right-handed charged currents
in the lepton sector. If we neglect the last type of currents we can
identify the $\xi$ parameter in Eq.~(\ref{p1}) in terms of the right-handed
mixing angles as follows:
\begin{equation}
\xi=\frac{\vert (V^{U\dagger}_R\,\Delta
V^D_R)_{cb}\vert}{\vert V_{cb}\vert}\approx 0.14\pm0.18.
\label{xinosso}
\end{equation}
Hence, we see that there is room for such processes in this quasi-chiral
model. There are $b$-decays induced by charged scalars too.

It is interesting to call attention to the fact that the measurement of the 
$\Lambda_b$ polarization gives
$P_\Lambda=-0.23^{+0.24}_{-0.20}({\rm stat.})^{+0.08}_{-0.07}({\rm syst.})$
~\cite{aleph}, while the value expected in the ESM is $P_\Lambda=-0.69\pm0.06$. 
(If right-handed currents do not exist this small polarization value
indicates that there are new mechanisms for depolarization which still have to 
be understood.)

As it has been emphasized by Gronau and Wakaizumi~\cite{lr} right-handed 
couplings of the $b$ quark to the $c$ and $u$ quarks would have effect in 
nonleptonic $b$ decays, extra contributions in $B^0_d-\bar{B}^0_d$ mixing,
in the $K_L-K_S$ mass difference and in $CP$ violation.
The study of $CP$ violation in $K$ and $B$ systems in the case of purely 
right-handed chirality for both $b$ to $c$ and 
$b$ to $u$ couplings has been done recently in Ref.~\cite{hayashi}.

In the previous discussion we have not considered the violation of the symmetry 
under $CP$. In the ESM we have in the charged current the mixing matrix 
$V_{\rm CKM}=V^{U\dagger}_LV^D_L$ which is a unitary $n\times n$ 
matrix in the case of $n$ quark generations.
It has $n^2-n(n-1)/2$ phases after the unitarity conditions have been
taking into account. On the other hand, even if the quarks are already the
mass eigenstates we have freedom for absorbing a phase into each left-handed
field: $u_{\alpha L}\to e^{if(u_\alpha)}u_{\alpha L}$ and 
$d_{\alpha L}\to e^{if(d_\alpha)}d_{\alpha L}$ (where $u_{\alpha L}=
u_L,c_L,t_L,...;\;\;{\rm and}\;\; d_{\alpha L}=
d_L,s_L,b_L,...$) which 
removes the arbitrary phase from one row or column of $V_{\rm CKM}$. But as 
this matrix is unchanged by a common phase transformation of all $u_{iL}$, only 
$2n-1$ phase degrees of freedom can be removed in this way. Therefore the
$V_{\rm CKM}$ matrix has only $(n-2)(n-1)/2$ physically independent phases.
This counting of the physical phases is possible only in the context of the 
ESM: the new phases in the left-handed fields are absorbed into the right-
handed fields in the mass term and in the neutral current
$u_{\alpha R}\to e^{if(u_\alpha)}u_{\alpha R}$ (similarly in the $d$-like 
right-handed components). 
Since these neutral currents are helicity and flavor conserving and the mass 
matrices have been already diagonalized, only $(n-2)(n-1)/2$ phases survive 
in the charged current. The other phases (and right-handed mixing matrices) 
will not appear at all in any place of the Lagrangian density. 

On the other hand, in quasi-chiral models, as the present one, there are 
nondiagonal right-handed currents in both, charged and neutral currents.
So we have no freedom to redefine the right-handed fields.
Hence, the number of phases 
in the $V_{\rm CKM}$ matrix is now $n^2-n(n-1)/2-1$ since we have still  
freedom for choosing a common phase in all left- and right-handed fields. 
For $n=3$ we have 4 phases instead of only one as in the ESM. 
We of course, can still insist in absorbing
the phases in $V_{\rm CKM}$ but the extra phases will appear in other places
of the Lagrangian density. We have in this case for instance
\begin{mathletters}
\label{ult}
\begin{equation}
V^{U\dagger}_L\Omega V^U_R\to K^{u\dagger}V^{U\dagger}_L\Omega V^U_RK^u,
\;\;
V^{D\dagger}_L\Omega V^D_R
\to 
K^{d\dagger}V^{D\dagger}_L\Omega V^D_RK^d,\;\;
V^{D\dagger}_L\Delta V^U_R\to K^{d\dagger}V^{D\dagger}_L\Delta V^U_RK^u,
\label{ult1}
\end{equation}
in Eq.~(\ref{yukawa}) and
\begin{equation}
V^{D\dagger}_R \Delta V^D_R\to K^{d\dagger}V^{D\dagger}_R \Delta V^D_RK^d,
\;\;V^{U\dagger}_R\Delta V^U_R\to
K^{u\dagger}V^{U\dagger}_R\Delta V^U_RK^u,
\label{ult2}
\end{equation}
in Eqs.~(\ref{n1}) and (\ref{n2}), respectively; also
\begin{equation}
V^{U\dagger}_R \Delta V^D_R\to
K^{u\dagger}V^{U\dagger}_R \Delta V^D_RK^u,
\label{ult3}
\end{equation}
in Eq.~(\ref{ru}), we have defined
\begin{equation}
K^u={\rm diag}(e^{if(u_1)},e^{if(u_2)},e^{if(u_3)}),
\quad K^d={\rm diag}(e^{if(d_1)},e^{if(d_2)},e^{if(d_3)}).
\label{ks}
\end{equation}
\end{mathletters}
This makes richer the 
phenomenology of the present model. For instance, there are new phases in
the non-diagonal neutral currents coupled to $Z^0$ as it is evident from Eqs.
(\ref{n1}), (\ref{n2}) and (\ref{ult2}) and also in the right-handed current 
as can be seen from Eqs.~(\ref{ru}) and (\ref{ult3}).

Summarizing, we have shown that it is possible to have (non-dominant) 
right-handed current effects in the context of a $SU(2)\otimes U(1)_Y$ model. 
These effects depend on the parameters $V^{U,D,l}_{L,R}$ which do not survive 
as observable parameters in the 
model when all generations transform in the same way i.e., as left-handed 
doublets and right-handed singlets. 

Notwithstanding, we recall that the possibility that all the three families 
transform in a vector-like way is inconsistent with the present data. The 
charged currents 
given in Eq.~(\ref{ru}) may still be phenomenological consistent since they 
involve the right-handed Cabibbo-like mixing matrix 
$V^{U\dagger}_R\Delta V^D_R$, then the dominant $V-A$ character of the 
charged weak interactions implies constraints only on these matrix 
elements. However, in this case the right-handed neutral current couplings, 
see Eqs.~(\ref{n1}) and (\ref{n2}), are 
all proportional to $R_Q-1,\;\;Q=U,D$, and this is in fact ruled out by 
currently $Z$-pole observables. On the other hand, it is still possible that 
two families transform in a vector-like way.
The $\Delta$ in Eqs.(\ref{n1})--(\ref{ru}) is in this case 
$\Delta\to \Delta'=diag(0,1,1)$. Of course, if we naively assume from the 
beginning that $V^{U,D,l}_R\approx{\bf1}$ the model is not consistent with 
data. It is necessary to assume that all these right-handed mixing matrices 
must be determined only by experiments.

We would like to stress once more that, all parameters appearing in the 
present model do exist in the electroweak standard model (here we have 
called it ESM) but the GIM mechanism~\cite{sm} cancels out all of them in the 
Lagrangian density except three angles and one phase. Hence, here we have not 
introduced new parameters but we only have shown that in the quasi-chiral 
model (and this will occur also in other extensions of the ESM with flavor 
changing neutral currents) some of the parameters of
the ESM do survive when we write the Lagrangian density in terms of the 
physical fields and even after absorbing all phases allowed by the interactions.

\acknowledgements
I am grateful to J. C. Montero for useful discussions and the Conselho 
Nacional de Desenvolvimento Cient\'\i fico e Tecnol\'ogico (CNPq) for partial 
financial support.


\begin{references}
\bibitem{sm} S. Weinberg, Phys. Rev. Lett. {\bf19},
(1967), 1264; A. Salam, in {\sl Elementary Particle Theory}, edited by N.
Svartholm (Almquist and Wiksell, 1968), p. 367; S. L. Glashow, Nucl. Phys.
 {\bf22} (1961), 579; S. L. Glashow, J. Iliopoulos and L. Maiani, Phys. Rev.
D {\bf2} (1970), 1285; M. Kobayashi and T. Maskawa, Prog. Theor. Phys. 
{\bf49} (1974), 652; M. Y. Han and Y. Nambu, Phys. Rev. (1965), B1006;
H. Fritzsch, M. Gell-Mann and H. Leutwyler, Phys. Lett. {\bf 47B} (1973), 365;
D. J. Gross and F. Wilczek, Phys. Rev. {\bf D8} (1973), 3633.
\bibitem{pdg} R. M. Barnett {\it et al.}, (Review of Particle Physics),
Phys. Rev. D {\bf54} (1996), 1.
\bibitem{new2} S. Dawson, DPF'96 The triumph of the standard model, 
hep-ph/9609340.
\bibitem{lr} M. Gronau and S. Wakaizumi, Phys. Rev. Lett. {\bf68}
(1992), 1814; Phys. Rev. D {\bf47} (1993), 1262; J. F. Amudson, J. L. Rosner, 
M. Worash and M. B. Wise, Phys. Rev. D {\bf 47} (1993), 1260; R. N. Mohapatra
and S. Nussinov, Phys. Lett. {\bf B339} (1994), 101. 
\bibitem{japa} K. Fujikawa and A. Yamada, Phys. Rev. D\ {\bf 49} (1994), 5890.
\bibitem{lrm} For early L-R symmetric model's references  see J. C. Pati 
and A. Salam, Phys. Rev. {\bf D10} (1974), 275; R. N. Mohapatra and J. C. 
Pati, Phys. Rev. {\bf D11} (1975), 566, 2558; G. Senjanovic and R. N. 
Mohapatra, Phys. Rev. {\bf D12} (1975), 1502; A. De R\'ujula, H. Georgi and 
S. L. Glashow, Ann. Phys. (N.Y.) {\bf109} (1977), 242;
R. N. Mohapatra and R. E. 
Marshak, Phys. Lett. {\bf 91B} (1980), 222. 
\bibitem{kuz} I. A. Kuznatson {\it et al}., Phys. Rev. Lett. {\bf75}
(1995), 794.
\bibitem{gb} G. Barbiellini and C. Santoni, R. Nuovo Cimento {\bf 9}(2) (1986), 
1.
\bibitem{gp} H. Georgi and A. Pais, Phys. Rev. D {\bf19} (1979), 2746.
\bibitem{vp} V. Pleitez, Phys. Rev. D {\bf53} (1996), 514.
\bibitem{ema} E. Ma, Phys. Rev. D\ {\bf53} (1996), R2276.
\bibitem{tau3} C. O. Escobar, O. L. G. Peres, V. Pleitez and R. Zukanovich
Funchal, Phys. Rev. D {\bf 47} (1993), R1747. For the general case of $n$ 
families of quarks and leptons see C. Jarlskog, Phys. Lett. {\bf B241} (1990), 
579. 
\bibitem{gim} S. L. Glashow, J. Iliopoulos and L. Maiani in Ref.~\cite{sm}.
\bibitem{cleo} S. Sanghera {\it et al.} (CLEO Collaboration), Phys. Rev. D
{\bf47} (1993), 791.
\bibitem{aa} H. Albrecht {\sl et al.} (ARGUS Collaboration), Phys. Lett. 
{\bf B250} (1990), 164; D. Buskulic {\sl et al.} (ALEPH Collaboration),
Phys. Lett. {\bf B321} (1994), 168.
\bibitem{aam} H. Albreght {\sl et al.} (ARGUS Collaboration), Phys. Lett. 
{\bf B246} (1990), 278;  {\bf B316} (1993), 608 and {\bf B341} (1995), 441; 
D. Buskulic {\sl et al.} (ALEPH Collaboration), Phy. Lett. {\bf B346} (1995), 
379.
\bibitem{cpb} P. Bamert, C. P. Burgess and I. Maksymyk, Phys. Lett. {\bf B356}
(1995), 282. 
\bibitem{charm} H. Abramowics {\it et al.}, Z. Phys. {\bf C15} (1982), 19.
\bibitem{L3} M. Acciari {\it et al.} (L3 Collaboration), Phys. Lett. {\bf B351} 
(1995, 375.
\bibitem{voloshin} M. B. Voloshin, Mod. Phys. Lett. {\bf A12} (1997), 1823.
\bibitem{aleph} D. Buskulic {\it et al.} (ALEPH Collaboration),
Phys. Lett. {\bf B365} (1996), 437.
\bibitem{hayashi} T. Hayashi, Prog. Theor. Phys. {\bf98} (1997), 143.
\end{references}
\end{document}